\def\versionstring{}

\documentclass[10pt,conference]{IEEEtran}

\usepackage{amssymb, stmaryrd}
\usepackage{amsfonts}
\usepackage{amsmath}
\usepackage{latexsym}
\usepackage{epsfig}
\usepackage{psfrag}
\usepackage{bm}
\usepackage{xspace}
\usepackage{graphicx}
\usepackage{subfigure}

\usepackage{watermark}

\IEEEtriggeratref{17}

\newlength{\watermarklength}
\newlength{\watermarklengthtwo}
\newlength{\test}
\watermark{\put(0,30){\versionstring}}

\begin{document}

\title{Interior-Point Algorithms for \\
       Linear-Programming Decoding${}^{0}$}

\author{\authorblockN{Pascal O.~Vontobel}
\authorblockA{Hewlett--Packard Laboratories\\
              Palo Alto, CA 94304, USA\\
              Email: pascal.vontobel@ieee.org}
}

\maketitle

\newcommand{\Expec}{\operatorname{E}}
\newcommand{\Var}{\operatorname{Var}}
\newcommand{\supp}{\operatorname{supp}}
\newcommand{\suml}{\sum\limits}
\newcommand{\sumsupp}[2]{\suml_{#1\in\supp(P_{#2}(.))}}
\newcommand{\prodl}{\prod\limits}
\newcommand{\intl}{\int\limits}
\newcommand{\Prob}{\operatorname{Pr}}
\newcommand{\dint}[1]{\operatorname{d}{#1}}
\newcommand{\alphaML}{\hat \alpha_{\mathrm{ML}}}
\newcommand{\argmax}[1]{\underset{#1}{\operatorname{argmax}}}
\newcommand{\dmins}{d_{\mathrm{min}}}
\newcommand{\matr}[1]{\mathbf{#1}}
\newcommand{\vect}[1]{\mathbf{#1}}
\newcommand{\code}[1]{\mathcal{#1}}
\newcommand{\ocodeA}{\overline{\code{A}}}
\newcommand{\ocodeB}{\overline{\code{B}}}
\newcommand{\set}[1]{\mathcal{#1}}
\newcommand{\graph}[1]{\mathsf{#1}}
\newcommand{\tree}[1]{\mathcal{#1}}
\newcommand{\forest}[1]{\mathcal{#1}}
\newcommand{\argmin}{\operatorname{arg min}}
\newcommand{\GF}[1]{\mathbb{F}_{#1}}
\newcommand{\R}{\mathbb{R}}
\newcommand{\Rp}{\mathbb{R}_{+}}
\newcommand{\Rpp}{\mathbb{R}_{++}}
\newcommand{\card}[1]{\lvert #1 \rvert}

\newcommand{\vecte}{\vect{e}}
\newcommand{\vecteV}{\vect{e}_{\mathrm{V}}}
\newcommand{\vecteC}{\vect{e}_{\mathrm{C}}}
\newcommand{\vecteR}{\vect{e}_{\mathrm{R}}}
\newcommand{\vecti}{\vect{i}}
\newcommand{\vectiV}{\vect{i}_{\mathrm{V}}}
\newcommand{\vectiC}{\vect{i}_{\mathrm{C}}}
\newcommand{\vectiR}{\vect{i}_{\mathrm{R}}}
\newcommand{\matN}{\matr{N}}
\newcommand{\matNV}{\matr{N}_{\mathrm{V}}}
\newcommand{\matNC}{\matr{N}_{\mathrm{C}}}
\newcommand{\matNR}{\matr{N}_{\mathrm{R}}}

\newcommand{\Uq}{I_{\mathrm{q}}}
\newcommand{\Iq}{I_{\mathrm{q}}}
\newcommand{\PG}[2]{\operatorname{PG}(#1,#2)}
\newcommand{\EG}[2]{\operatorname{EG}(#1,#2)}
\newcommand{\EGq}{\operatorname{EG}(2,q)}
\newcommand{\PGq}{\operatorname{PG}(2,q)}
\newcommand{\codePGq}{\code{C}_{\PGq}}
\newcommand{\codeEGq}{\code{C}_{\EGq}}
\newcommand{\wcol}{w_{\mathrm{col}}}
\newcommand{\wrow}{w_{\mathrm{row}}}
\newcommand{\defeq}{\triangleq}
\newcommand{\setDML}[1]{\set{D}^{\mathrm{ML}}_{#1}}
\newcommand{\setDMLzero}{\set{D}^{\mathrm{ML}}_{\vect{0}}}
\newcommand{\setDLPzero}{\set{D}^{\mathrm{LP}}_{\vect{0}}}

\newcommand{\sonenorm}[1]{\lVert #1 \rVert_1}
\newcommand{\onenorm}[1]{\lVert #1 \rVert_1}
\newcommand{\twonorm}[1]{\lVert #1 \rVert_2}
\newcommand{\infnorm}[1]{\lVert #1 \rVert_{\infty}}

\newcommand{\FBethe}{F_{\mathrm{Bethe}}}

\newtheorem{Definition}{Definition}
\newtheorem{Example}[Definition]{Example}
\newtheorem{Assumption}[Definition]{Assumption}
\newtheorem{Comment}[Definition]{Comment}
\newtheorem{Remark}[Definition]{Remark}

\newtheorem{Lemma}[Definition]{Lemma}
\newtheorem{Theorem}[Definition]{Theorem}
\newtheorem{Proposition}[Definition]{Proposition}
\newtheorem{Corollary}[Definition]{Corollary}
\newtheorem{Algorithm}[Definition]{Algorithm}
\newtheorem{Conjecture}[Definition]{Conjecture}

\newenvironment{Proof}%
  {\noindent \emph{Proof:}}{\hfill$\square$}

\newcommand{\eclaim}{\hfill$\square$}
\newcommand{\eproposition}{\hfill$\square$}
\newcommand{\elemma}{\hfill$\square$}
\newcommand{\etheorem}{\hfill$\square$}
\newcommand{\edefinition}{\mbox{ }\hfill$\square$}
\newcommand{\econjecture}{\hfill$\square$}
\newcommand{\fedefinition}{\hfill\square}
\newcommand{\eexample}{\hfill$\square$}
\newcommand{\eassumption}{\hfill$\square$}
\newcommand{\ecorollary}{\hfill$\square$}
\newcommand{\equestion}{\hfill$\square$}
\newcommand{\efact}{\hfill$\square$}
\newcommand{\eremark}{\hfill$\square$}
\newcommand{\RM}{\operatorname{RM}}
\newcommand{\convhull}{\operatorname{conv}}
\newcommand{\conichull}{\operatorname{conic}}

\newcommand{\rank}{\operatorname{rank}}

\newcommand{\wps}{w_{\mathrm{p}}}
\newcommand{\wpsAWGNC}{w_{\mathrm{p}}^{\mathrm{AWGNC}}}
\newcommand{\wpsBSC}{w_{\mathrm{p}}^{\mathrm{BSC}}}
\newcommand{\wpsBEC}{w_{\mathrm{p}}^{\mathrm{BEC}}}
\newcommand{\wpsmin}{w_{\mathrm{p}}^{\mathrm{min}}}
\newcommand{\wpsminc}[1]{\wpsmin(\code{#1})}
\newcommand{\wpsminh}[1]{\wpsmin(\matr{#1})}
\newcommand{\wpsBSCmin}{w_{\mathrm{p}}^{\mathrm{BSC,min}}}
\newcommand{\wpsBECmin}{w_{\mathrm{p}}^{\mathrm{BEC,min}}}

\newcommand{\type}{t}
\newcommand{\vtype}{\vect{t}}

\newcommand{\va}{\vect{a}}
\newcommand{\tva}{\tilde{\vect{a}}}
\newcommand{\vb}{\vect{b}}
\newcommand{\tvb}{\tilde{\vect{b}}}
\newcommand{\vc}{\vect{c}}
\newcommand{\tvc}{\tilde{\vect{c}}}
\newcommand{\hvc}{\vect{\hat c}}
\newcommand{\htvc}{\vect{\hat {\tilde c}}}
\newcommand{\vd}{\vect{d}}
\newcommand{\tvd}{\tilde{\vect{d}}}
\newcommand{\ve}{\vect{e}}
\newcommand{\vf}{\vect{f}}
\newcommand{\vg}{\vect{g}}
\newcommand{\vgs}{\vect{g}^{*}}
\newcommand{\vh}{\vect{h}}
\newcommand{\vm}{\vect{m}}
\newcommand{\vms}{\vect{m}^{*}}
\newcommand{\vu}{\vect{u}}
\newcommand{\vus}{\vect{u}^{*}}
\newcommand{\tvu}{\tilde{\vect{u}}}
\newcommand{\ovu}{\overline{\vect{u}}}
\newcommand{\vv}{\vect{v}}
\newcommand{\tvv}{\tilde{\vect{v}}}
\newcommand{\uvv}{\vect{\breve{v}}}
\newcommand{\vw}{\vect{w}}
\newcommand{\vws}{\vect{w}^{*}}
\newcommand{\vX}{\vect{X}}
\newcommand{\vx}{\vect{x}}
\newcommand{\vhx}{\vect{\hat x}}
\newcommand{\vxs}{\vect{x}^{*}}
\newcommand{\vY}{\vect{Y}}
\newcommand{\vy}{\vect{y}}
\newcommand{\vgamma}{\boldsymbol{\gamma}}
\newcommand{\vdelta}{\boldsymbol{\delta}}
\newcommand{\vlambda}{\boldsymbol{\lambda}}
\newcommand{\tvlambda}{\tilde{\boldsymbol{\lambda}}}
\newcommand{\vmu}{\boldsymbol{\mu}}
\newcommand{\tvmu}{\tilde{\boldsymbol{\mu}}}
\newcommand{\vtau}{\boldsymbol{\tau}}
\newcommand{\ovtau}{\overline{\vtau}}
\newcommand{\oovtau}{\overline{\overline{\vtau}}}
\newcommand{\vxi}{\boldsymbol{\xi}}

\newcommand{\vomega}{\boldsymbol{\omega}}
\newcommand{\bvomega}{\bar{\boldsymbol{\omega}}}
\newcommand{\hvomega}{\hat{\boldsymbol{\omega}}}
\newcommand{\vGamma}{\boldsymbol{\Gamma}}
\newcommand{\bo}{\boldsymbol{1}}
\newcommand{\bz}{\boldsymbol{0}}

\newcommand{\vzero}{\vect{0}}

\newcommand{\setoP}{\overline{\set{P}}}
\newcommand{\setoD}{\overline{\set{D}}}

\newcommand{\PLP}{P_{\mathrm{LP}}}
\newcommand{\DLP}{D_{\mathrm{LP}}}
\newcommand{\PQP}{P_{\mathrm{QP}}}
\newcommand{\DQP}{D_{\mathrm{QP}}}
\newcommand{\PQPeps}{P_{\mathrm{QP}}(\varepsilon)}
\newcommand{\DQPeps}{D_{\mathrm{QP}}(\varepsilon)}

\newcommand{\IP}{\operatorname{IP}}
\newcommand{\LP}{\operatorname{LP}}
\newcommand{\QP}{\operatorname{QP}}
\newcommand{\QPeps}{\operatorname{QP}(\varepsilon)}

\newcommand{\tr}{\mathsf{T}}

\newcommand{\Eb}{E_{\mathrm{b}}}
\newcommand{\Ec}{E_{\mathrm{c}}}

\newcommand{\fcheck}{f_{\mathrm{ch}}}
\newcommand{\IXOR}{I_{\mathrm{XOR}}}
\newcommand{\IPT}{I_{\mathrm{PT}}}
\newcommand{\ICHCode}[1]{I_{\mathrm{CH}(#1)}}
\newcommand{\ICHSPC}{I_{\mathrm{CHSPC}}}
\newcommand{\ICHRC}{I_{\mathrm{CHRC}}}
\newcommand{\LXOR}{L_{\mathrm{XOR}}}
\newcommand{\LPT}{L_{\mathrm{PT}}}
\newcommand{\LCHCode}[1]{L_{\mathrm{CH}(#1)}}
\newcommand{\LCHSPC}{L_{\mathrm{CHSPC}}}
\newcommand{\LCHRC}{L_{\mathrm{CHRC}}}
\newcommand{\fIP}{f_{\mathrm{IP}}}
\newcommand{\fLP}{f_{\mathrm{LP}}}
\newcommand{\fLPsym}{f_{\mathrm{LP}}^{\mathrm{sym}}}
\newcommand{\tfLP}{\tilde f_{\mathrm{LP}}}
\newcommand{\fQP}{f_{\mathrm{QP}}}
\newcommand{\fQPeps}{f_{\mathrm{QP(\varepsilon)}}}
\newcommand{\tfQP}{\tilde f_{\mathrm{QP}}}
\newcommand{\tfQPeps}{\tilde f_{\mathrm{QP(\varepsilon)}}}

\newcommand{\Niter}{N_{\mathrm{iter}}}

\newcommand{\Sconvexhull}{\set{S}_{\mathrm{conv.~hull}}}
\newcommand{\Srelax}{\set{S}_{\mathrm{relax}}}

\newcommand{\exampleend}{\hfill$\square$}

\newcommand{\fp}[1]{\mathcal{#1}}
\newcommand{\pfp}[1]{\dot{\mathcal{#1}}}
\newcommand{\fc}[1]{\mathcal{#1}}
\newcommand{\pfc}[1]{\dot{\mathcal{#1}}}
\newcommand{\fph}[2]{\mathcal{#1}(\matr{#2})}
\newcommand{\pfph}[2]{\dot{\mathcal{#1}}(\matr{#2})}
\newcommand{\fch}[2]{\mathcal{#1}(\matr{#2})}
\newcommand{\pfch}[2]{\dot{\mathcal{#1}}(\matr{#2})}

\newcommand{\vnu}{\boldsymbol{\nu}}
\newcommand{\vzeta}{\boldsymbol{\zeta}}

\newcommand{\matrunity}{\matr{I}}

\newcommand{\alg}[1]{\mathcal{#1}}
\newcommand{\MSA}{\mathrm{MSA}}
\newcommand{\IPFFG}{\mathrm{IP-FFG}}
\newcommand{\LPFFG}{\mathrm{LP-FFG}}
\newcommand{\QPepsFFG}{\mathrm{QP}(\varepsilon)\mathrm{-FFG}}

\newcommand{\Ra}{R_{\mathrm{a}}}
\newcommand{\Raz}{R_{\mathrm{a},0}}
\newcommand{\Rb}{R_{\mathrm{b}}}
\newcommand{\Rbz}{R_{\mathrm{b},0}}

\newcommand{\del}{\partial}

\newcommand{\e}{\mathrm{e}}

\newcommand{\gC}{\graph{C}}
\newcommand{\gB}{\graph{B}}

\newcommand{\btau}{{\boldsymbol{\tau}}}
\newcommand{\ctau}{\btau^{(c)}}
\newcommand{\vvtau}[1]{\tau^{(v,#1)}}
\newcommand{\cctau}[1]{\tau^{(c,#1)}}

\newcommand{\Gammapos}{\Gamma_{\mathrm{pos}}}
\newcommand{\gammapos}{\gamma_{\mathrm{pos}}}
\newcommand{\Gammaneg}{\Gamma_{\mathrm{neg}}}
\newcommand{\gammaneg}{\gamma_{\mathrm{neg}}}

\newcommand{\setIpos}{\set{I}_{\mathrm{pos}}}
\newcommand{\setIneg}{\set{I}_{\mathrm{neg}}}

\newcommand{\const}{\mathrm{const}}

\newcommand{\intextbinomial}[2]{\mbox{ \scriptsize $\!\!\left(\!\!
      \begin{array}{c}
        #1 \\
        #2
      \end{array} \!\!\right)\!$}}

\newcommand{\optprog}[2]
{%
  \noindent\mbox{}\\[0cm]
  \noindent\fbox{%
  \begin{minipage}{0.955\linewidth}
    \mbox{}\\[-0.5cm]
    #1\\[#2]
  \end{minipage}
  }
  \noindent\mbox{}\\[-0.2cm]
}

\newcommand{\vdlambda}{\boldsymbol{\Delta\lambda}}
\newcommand{\vds}{\boldsymbol{\Delta}\vect{s}}
\newcommand{\vdx}{\boldsymbol{\Delta}\vect{x}}

\newcommand{\matrXi}{\boldsymbol{\Xi}}
\newcommand{\diag}{\operatorname{diag}}

\newcommand{\etal}{\emph{et al.}}

\footnotetext{This is essentially the paper that appeared in the Proceedings
  of the 2008 Information Theory and Applications Workshop, UC San Diego, CA,
  USA, January 27 -- February 1, 2008. The only (major) change concerns the
  vector $\vgamma$: it is now defined such that $\vhx_{\mathrm{ML}}$,
  $\vhx'_{\mathrm{ML}}$, and $\vhx_{\mathrm{LP}}$ are solutions to
  \emph{minimization} (and not \emph{maximization}) problems.}

\begin{abstract}
  Interior-point algorithms constitute a very interesting class of algorithms
  for solving linear-programming problems. In this paper we study efficient
  implementations of such algorithms for solving the linear program that
  appears in the linear-programming decoder formulation.
\end{abstract}

\section{Introduction}

\label{sec:introduction:1}

Consider a binary linear code $\code{C}$ of length $n$ that is used for data
transmission over a binary-input discrete memoryless channel. As was observed
by Feldman \etal~\cite{Feldman:03:1, Feldman:Wainwright:Karger:05:1}, the ML
decoder for this setup can be written as
\begin{align*}
  \vhx_{\mathrm{ML}}
    &= \arg \min_{\vx \in \code{C}} \
         \langle \vgamma, \vx \rangle,
\end{align*}
where $\vgamma$ is a length-$n$ vector that contains the log-likelihood ratios
and where $\langle \vgamma, \vx \rangle$ is the inner product (in $\R$) of the
vector $\vgamma$ with the vector $\vx$. Because the cost function in this
minimization problem is linear, this is essentially equivalent to the solution
of
\begin{align*}
  \vhx'_{\mathrm{ML}}
    &\defeq
       \arg \min_{\vx \in \convhull(\code{C})}
         \langle \vgamma, \vx \rangle,
\end{align*}
where $\convhull(\code{C})$ denotes the convex hull of $\code{C}$ when
$\code{C}$ is embedded in $\R^n$. (We say ``essentially equivalent'' because
in the case where there is a unique optimal codeword then the two minimization
problems yield the same solution. However, when there are multiple optimal
codewords then $\vhx_{\mathrm{ML}}$ and $\vhx'_{\mathrm{ML}}$ are non-singlet
sets and it holds that $\convhull(\vhx_{\mathrm{ML}}) = \vhx'_{\mathrm{ML}}$.)

Because the above two optimization problems are usually practically
intractable, Feldman \etal~\cite{Feldman:03:1,
  Feldman:Wainwright:Karger:05:1} proposed to solve a relaxation of the above
problem. Namely, for a code $\code{C}$ that can be written as the intersection
of $m$ binary linear codes of length $n$, i.e., $\code{C} \defeq
\cap_{j=1}^{m} \code{C}_j$, they introduced the so-called linear programming
(LP) decoder
\begin{align}
  \vhx_{\mathrm{LP}}
    &\defeq
       \arg \min_{\vx \in \set{P}} \
         \langle \vgamma, \vx \rangle,
           \label{eq:lp:decoder:1}
\end{align}
with the relaxed polytope
\begin{align}
  \set{P}
    &\defeq
       \bigcap_{j=1}^{m} \convhull(\code{C}_j)
     \supseteq
       \convhull(\code{C})
     \supseteq
       \code{C},
         \label{eq:fundamental:polytope:1}
\end{align}
for which it can easily be shown that all codewords in $\code{C}$ are vertices
of $\set{P}$.

The same polytope $\set{P}$ appeared also in papers by Koetter and Vontobel
~\cite{Koetter:Vontobel:03:1, Vontobel:Koetter:05:1:subm,
  Vontobel:Koetter:04:2}, where message-passing iterative (MPI) decoders were
analyzed and where this polytope $\set{P}$ was called the fundamental
polytope. The appearance of the same object in these two different contexts
suggests that there is a tight connection between LP decoding and MPI
decoding.

The above codes $\code{C}_j$ can be any codes of length $n$, however, in the
following we will focus on the case where these codes are codes of dimension
$n-1$. For example, let $\matr{H}$ be an $m \times n$ parity-check matrix for
the code $\code{C}$ and let $\vect{h}_j^\tr$ be the $j$-th row of
$\matr{H}$.\footnote{Note that in this paper all vectors are \emph{column}
  vectors.}  Then, defining
\begin{align*}
  \code{C}_j
    &= \big\{
         \vx \in \{ 0, 1 \}^n
         \ \big| \
         \langle \vect{h}_j, \vx \rangle = 0 \text{ (mod $2$)}
       \big\}
\end{align*}
for $j = 1, \ldots, m$, we obtain $\code{C} = \cap_{j=1}^{m} \code{C}_j$.

Of course, the reason why the decoder in~\eqref{eq:lp:decoder:1} is called LP
decoder is because the optimization problem in that equation is a linear
program (LP).\footnote{We use LP to denote both ``linear programming'' and
  ``linear program.''} There are two standard forms for LPs, namely
\begin{align}
  \text{minimize}\quad
    &\langle \vect{c}, \vect{x} \rangle
       \nonumber \\
  \text{subj.\ to}\quad
    &\matr{A} \vect{x} = \vect{b}
       \label{eq:lp:first:standard:form:1} \\
    &\vect{x} \geq \vect{0}
       \nonumber
\end{align}
and
\begin{align}
  \text{maximize}\quad
    &\langle \vect{b}, \vlambda \rangle
       \nonumber \\
  \text{subj.\ to}\quad
    &\matr{A\!}^\tr \vlambda + \vect{s} = \vect{c} 
       \label{eq:lp:second:standard:form:1} \\
    &\vect{s} \geq \vect{0}
       \nonumber
\end{align}
Any LP can be reformulated (by introducing suitable auxiliary variables, by
reformulating equalities as two inequalities, etc.) so that it looks like the
first standard form. Any LP can also be reformulated so that it looks like the
second standard form. Moreover, the first and second standard form are tightly
related in the sense that they are dual convex programs.  Usually, the LP
in~\eqref{eq:lp:first:standard:form:1} is called the primal LP and the LP
in~\eqref{eq:lp:second:standard:form:1} is called the dual LP. (As it is to be
expected from the expression ``duality,'' the primal LP is the dual of the
dual LP.)

Not unexpectedly, there are many ways to express the LP that appears
in~\eqref{eq:lp:decoder:1} in either the first or the second standard form,
and each of these reformulations has its advantages (and disadvantages). Once
it is expressed in one of the standard forms, any general-purpose LP solver
can basically be used to obtain the LP decoder output. However, the LP at hand
has a lot of structure and one should take advantage of it in order to obtain
very fast algorithms that can compete complexity- and time-wise with MPI
decoders.

Several ideas have been presented in the past in this direction, e.g., by
Feldman \etal~\cite{Feldman:Karger:Wainwright:02:1} who briefly mentioned the
use of sub-gradient methods for solving the LP of an early version of the LP
decoder (namely for turbo-like codes), by Yang et
al.~\cite{Yang:Wang:Feldman:05:1, Yang:Wang:Feldman:07:2} on efficiently
solvable variations of the LP decoder, by Taghavi and
Siegel~\cite{Taghavi:Siegel:06:1} on cutting-hyperplane-type approaches, by
Vontobel and Koetter~\cite{Vontobel:Koetter:07:1} on coordinate-ascent-type
approaches, by Dimakis and Wainwright~\cite{Dimakis:Wainwright:06:1} and by
Draper \etal~\cite{Draper:Yedidia:Wang:07:1} on improvements upon the LP
decoder solution, and by Taghavi and Siegel~\cite{Taghavi:Siegel:07:1} and by
Wadayama~\cite{Wadayama:07:1:subm} on using variations of LP decoding
(together with efficient implementations) for intersymbol-interference
channels.

In this paper our focus will be on so-called interior-point algorithms, a type
of LP solvers that has become popular with the seminal work of
Karmarkar~\cite{Karmarkar:84:1}. (After the publication
of~\cite{Karmarkar:84:1} in 1984, earlier work on interior-point-type
algorithms by Dikin~\cite{Dikin:67:1} and others became more widely known). We
present some initial thoughts on how to use this class of algorithms in the
context of LP decoding. So far, with the notable exception
of~\cite{Wadayama:07:1:subm}, interior-point-type algorithms that are
especially targeted to the LP in~\eqref{eq:lp:decoder:1} do not seem to have
been considered. One of our goals by pursuing these type of methods is that we
can potentially obtain algorithms that are better analyzable than MPI
decoders, especially when it comes to finite-length codes.
(Wadayama~\cite{Wadayama:07:1:subm} discusses some efficient
interior-point-type methods, however, he is trying to minimize a quadratic
cost function, and the final solution is obtained through the use of the
sum-product algorithm that is initialized by the result of the interior-point
search. Although~\cite{Wadayama:07:1:subm} presents some very interesting
approaches that are worthwhile pursuing, it is not quite clear if these
algorithms are better analyzable than MPI decoders.)

There are some interesting facts about interior-point-type algorithms that
make them worthwhile study objects. First of all, there are variants for which
one can prove polynomial-time convergence (even in the worst case, which is in
contrast to the simplex algorithm). Secondly, we can round an intermediate
result to the next vector with only $0$ / $\frac{1}{2}$ / $1$ entries and
check if it is a codeword.\footnote{To be precise, by rounding we mean that
  coordinates below $\frac{1}{2}$ are mapped to $0$, that coordinates above
  $\frac{1}{2}$ are mapped to $1$, and that coordinates equal to $\frac{1}{2}$
  are mapped to $\frac{1}{2}$.} (This is very similar to the stopping
criterion that is used for MPI algorithms.) Note that a similar approach will
probably not work well for simplex-type algorithms that typically wander from
vertex to vertex of the fundamental polytope. The reason is that rounding the
coordinates of a vertex yields only a codeword if the vertex was a
codeword.\footnote{Proof: in an LDPC code where all checks have degree at
  least two, the largest coordinate of any nonzero-vector vertex is at least
  $\frac{1}{2}$. Therefore, there is no nonzero-vector vertex that is rounded
  to the all-zero codeword. The proof is finished by using the symmetry of the
  fundamental polytope, i.e., the fact that the fundamental polytope ``looks''
  the same from any codeword.}  Thirdly, interior-point-type algorithms are
also interesting because they are less sensitive than simplex-type algorithms
to degenerate vertices of the feasible region; this is important because the
fundamental polytope has many degenerate vertices.

The present paper is structured as follows. In
Secs.~\ref{sec:affine:scaling:algorithm:1}
and~\ref{sec:primal:dual:interior:point:algorithms:1} we discuss two classes
of interior-point algorithms, namely affine-scaling algorithms and primal-dual
interior-point algorithms, respectively. As we will see, the bottleneck step
of the algorithms in these two sections is to repeatedly find the solution to
a certain type of system of linear equations. Therefore, we will address this
issue, and efficient solutions to it, in
Sec.~\ref{sec:solving:equation:system:1}.  Finally, we briefly mention some
approaches for potential algorithm simplifications in
Sec.~\ref{sec:other:simplifications:1} and we conclude the paper in
Sec.~\ref{sec:conclusions:1}.

\section{Affine-Scaling Algorithms}

\label{sec:affine:scaling:algorithm:1}

\begin{figure}
  \begin{center}
    \subfigure[]{\epsfig{file=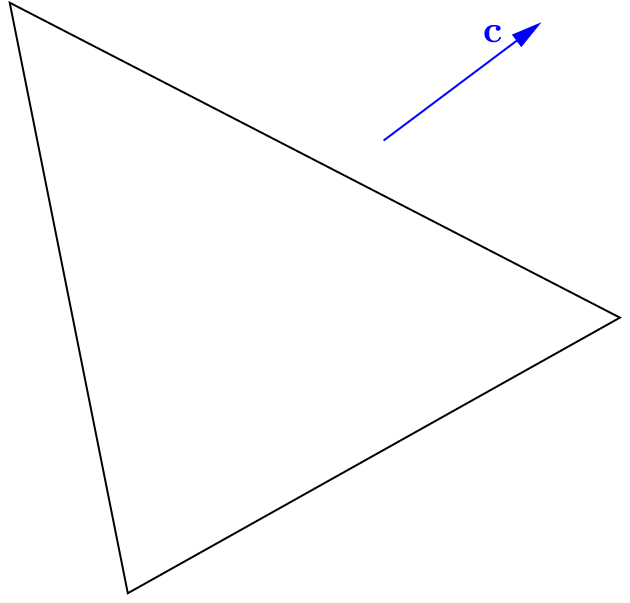, 
                  width=0.42\linewidth}}
    \subfigure[]{\epsfig{file=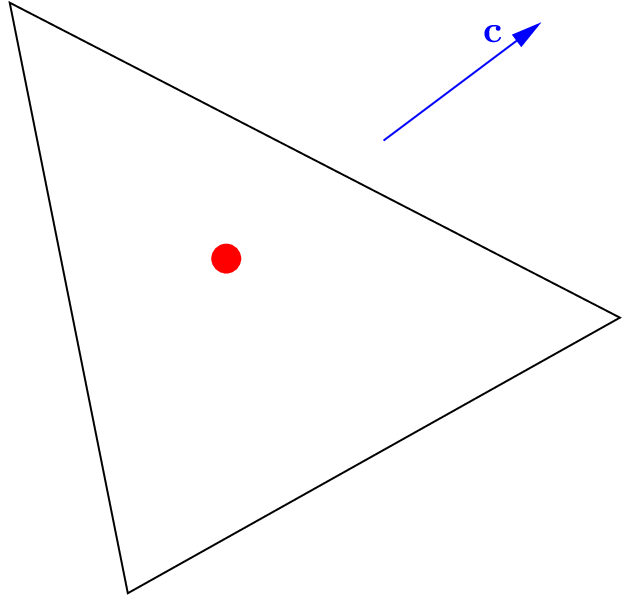, 
                  width=0.42\linewidth}}
    \subfigure[]{\epsfig{file=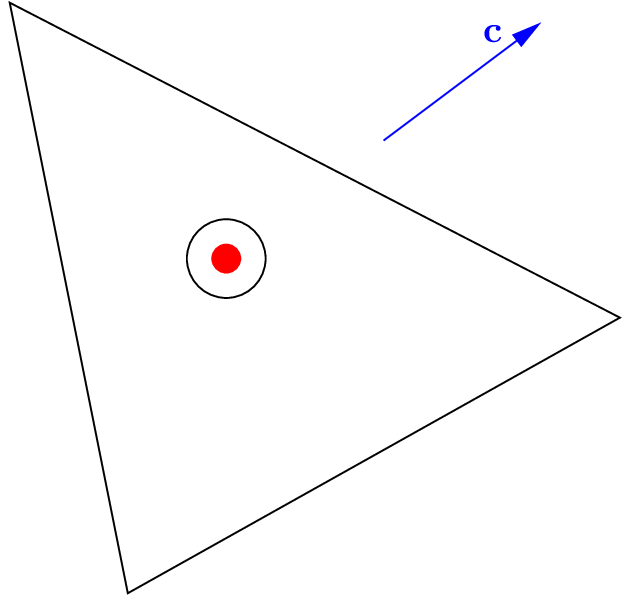, 
                  width=0.42\linewidth}}
    \subfigure[]{\epsfig{file=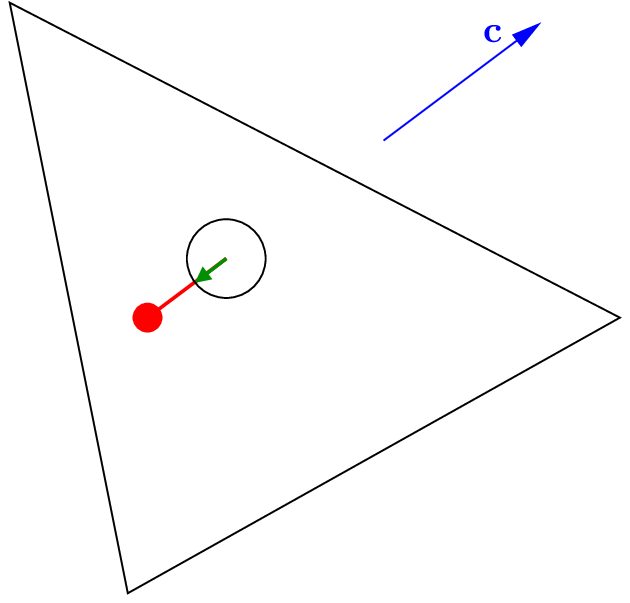, 
                  width=0.42\linewidth}}
    \subfigure[]{\epsfig{file=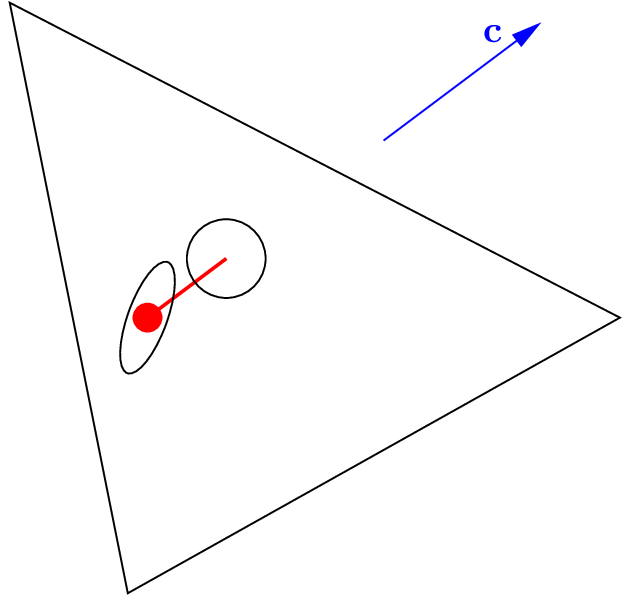, 
                  width=0.42\linewidth}}
    \subfigure[]{\epsfig{file=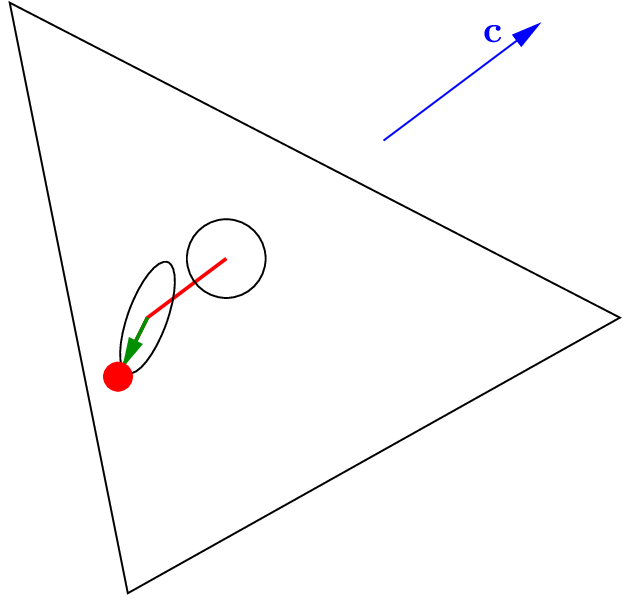, 
                  width=0.42\linewidth}}
    \subfigure[]{\epsfig{file=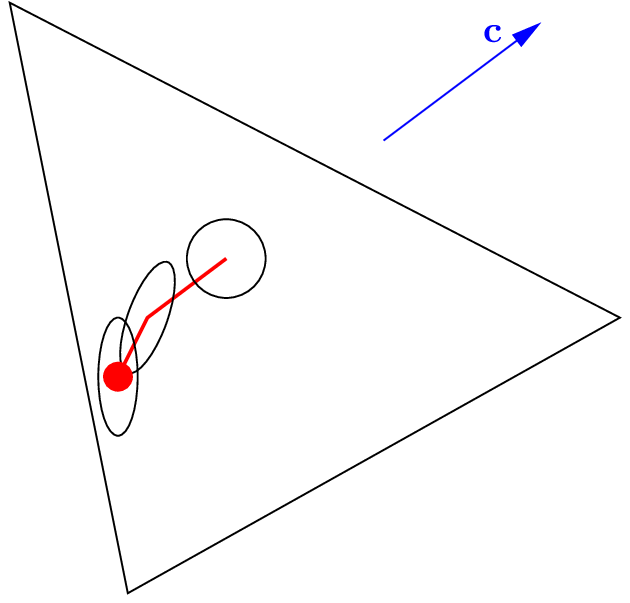, 
                  width=0.42\linewidth}}
    \subfigure[]{\epsfig{file=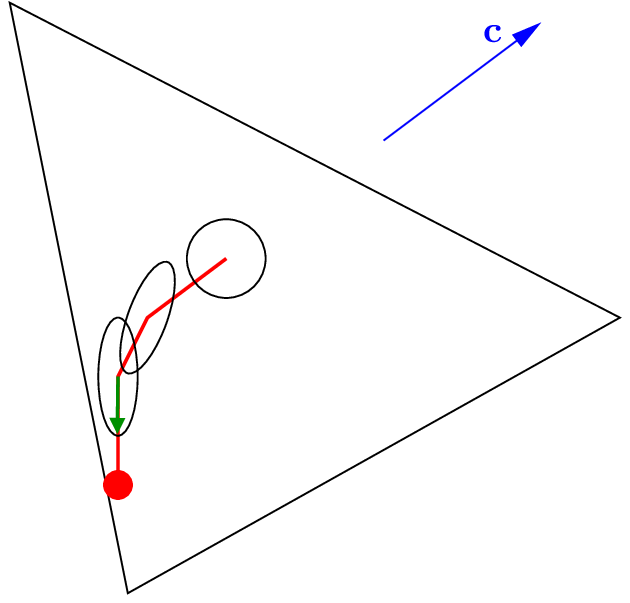, 
                  width=0.42\linewidth}}
    \subfigure[]{\epsfig{file=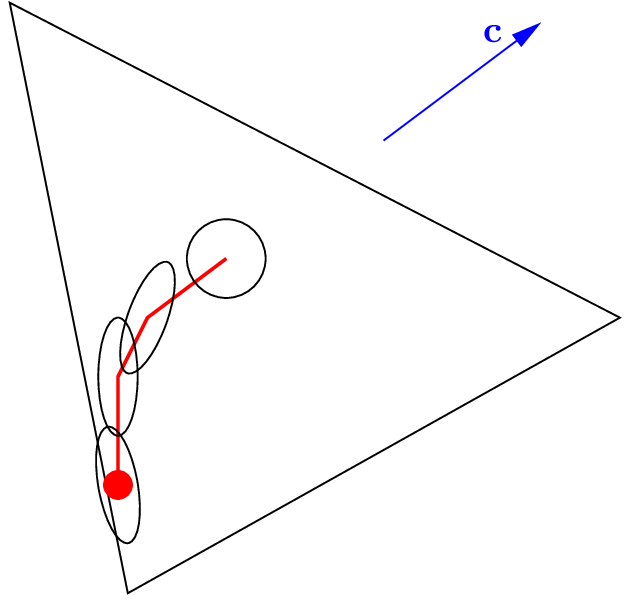, 
                  width=0.42\linewidth}}
    \subfigure[]{\epsfig{file=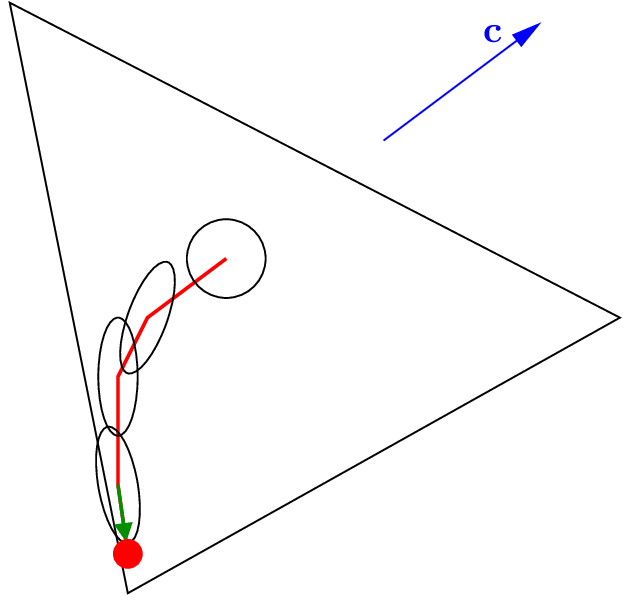, 
                  width=0.42\linewidth}}
  \end{center}
  \caption{Some iterations of the affine-scaling algorithm. (See text for
    details.)}
  \label{fig:affine:scaling:algorithm:iterations:1}
\end{figure}

An interesting class of interior-point-type algorithms are so-called affine
scaling algorithms which were introduced by Dikin~\cite{Dikin:67:1} and
re-invented many times afterwards. Good introductions to this class of
algorithms can be found in~\cite{Bertsekas:99:1, Bertsimas:Tsitsiklis:97:1}.

Fig.~\ref{fig:affine:scaling:algorithm:iterations:1} gives an intuitive
picture of the workings of one instance of an affine-scaling algorithm.
Consider the LP in~\eqref{eq:lp:first:standard:form:1} and assume that the set
of all feasible points, i.e., the set of all $\vx$ such that $\matr{A} \vx =
\vect{b}$ and $\vx \geq \vect{0}$, is a triangle. For the vector $\vect{c}$
shown in Fig.~\ref{fig:affine:scaling:algorithm:iterations:1}, the optimal
solution will be the vertex in the lower left part. The algorithm works as
follows:
\begin{enumerate}

\item Select an initial point that is in the interior of the set of all
  feasible points,
  cf.~Fig.~\ref{fig:affine:scaling:algorithm:iterations:1}(b), and let the
  current point be equal to this initial point.

\item Minimizing $\langle \vect{c}, \vect{x} \rangle $ over the triangle is
  difficult (in fact, it is the problem we are trying to solve); therefore, we
  replace the triangle constraint by an ellipsoidal constraint that is
  centered around the current point. Such an ellipsoid is shown in
  Fig.~\ref{fig:affine:scaling:algorithm:iterations:1}(c). Its skewness
  depends on the closeness to the different boundaries. 

\item We then minimize the function $\langle \vect{c}, \vect{x} \rangle$ over
  this ellipsoid. The difference vector between this minimizing point and the
  center of the ellipsoid (see the little vector in
  Fig.~\ref{fig:affine:scaling:algorithm:iterations:1}(d)) points in the
  direction in which the next step will be taken.

\item Depending on what strategy is pursued, a shorter or a longer step in the
  above-found direction is taken. This results in a new current point.
  (Whatever step size is taken, we always impose the constraint that the step
  size is such that the new current point lies in the interior of the set of
  feasible points.)

\item If the current point is ``close enough'' to some vertex then stop,
  otherwise go to Step 2. (``Closeness'' is determined according to some
  criterion.)

\end{enumerate}
Not surprisingly, when short (long) steps are taken in Step~4, the resulting
algorithm is called the short-step (long-step) affine-scaling algorithm.
Convergence proofs for different cases can be found in~\cite{Tsuchiya:91:1,
  Tsuchiya:Muramatsu:95:1, Saigal:96:1}.

Of course, an affine-scaling algorithm can also be formulated for the LP
in~\eqref{eq:lp:second:standard:form:1}. Moreover, instead of the
above-described discrete-time version, one can easily come up with a
continuous-time version, see e.g.~\cite{Adler:Monteiro:91:1}. The latter type
of algorithms might actually be interesting for decoders that are implemented
in analog VLSI.

The bottleneck step in the affine-scaling algorithm is to find the new
direction, which amounts to solving a system of linear equations of the form
$\matr{P} \vect{u} = \vect{v}$, where $\matr{P}$ is a given
(iteration-dependent) positive definite matrix, $\vect{v}$ is a given vector,
and $\vect{u}$ is the direction vector that needs to be found. We will comment
on efficient approaches for solving such systems of equations in
Sec.~\ref{sec:solving:equation:system:1}.

\section{Primal-Dual Interior Point Algorithms}

\label{sec:primal:dual:interior:point:algorithms:1}

In contrast to affine-scaling algorithms, which either work only with the
primal LP or only with the dual LP, primal-dual interior point algorithms --
as the name suggests -- work simultaneously on obtaining a primal \emph{and} a
dual optimal solution. A very readable and detailed introduction to this topic
can be found in~\cite{Wright:97:1}. As in the case of the affine-scaling
algorithm there are many different variations: short-step, long-step,
predictor-corrector, path-following, etc.

Again, the bottleneck step is to find the solution to a system of linear
equations $\matr{P} \vect{u} = \vect{v}$, where $\matr{P}$ is a given
(iteration-dependent) positive definite matrix, $\vect{v}$ is a given
(iteration-dependent) vector, and $\vect{u}$ is the sought quantity. We will
comment in Sec.~\ref{sec:solving:equation:system:1} on how such systems of
linear equations can be solved efficiently.

A variant that is worthwhile to be mentioned is the class of so-called
infeasible-interior-point algorithms. The reason is that very often it is easy
to find an initial primal feasible point or it is easy to find an initial dual
feasible point but not both at the same time. Therefore, one starts the
algorithm with a primal/dual point pair where the primal and/or the dual point
are infeasible points; the algorithm then tries to decrease the amount of
``infeasibility'' (a quantity that we will not define here) at every
iteration, besides of course optimizing the cost function.

One of the most intriguing aspects of primal-dual interior-point algorithms is
the polynomial-time worst-case bounds that can be stated. Of course, these
bounds say mostly something about the behavior when the algorithm is already
close to the solution vertex. It remains to be seen if these results are
useful for implementations of the LP decoder where it is desirable that the
initial iterations are as aggressive as possible and where the behavior close
to a vertex is not that crucial. (We remind the reader of the
rounding-procedure that was discussed at the end of
Sec.~\ref{sec:introduction:1}, a procedure that took advantage of some special
properties of the fundamental polytope.)

\section{Efficient Approaches for Solving 
             $\matr{P} \vect{u} = \vect{v}$ where
             $\matr{P}$ is a Positive Definite Matrix}

\label{sec:solving:equation:system:1}

In Secs.~\ref{sec:affine:scaling:algorithm:1}
and~\ref{sec:primal:dual:interior:point:algorithms:1} we saw that the crucial
part in the discussed algorithms was to repeatedly and efficiently solve a
system of linear equations that looks like
\begin{align*}
  \matr{P} \vect{u}
    &= \vect{v},
\end{align*}
where $\matr{P}$ is an iteration-dependent positive definite matrix and where
$\vect{v}$ is an iteration-dependent vector. The fact that $\matr{P}$ is
positive definite helps because $\vect{u}$ can also be seen to be the solution
of the quadratic unconstrained optimization problem
\begin{align}
  \text{minimize}\quad
    &\frac{1}{2}
     \vect{u}^\tr \matr{P} \vect{u}
     - \langle
         \vect{v}, \vect{u}
       \rangle
         \label{eq:unconstrained:qp:1} \\
  \text{subj.\ to}\quad
    &\vu \in \R^h
       \nonumber
\end{align}
where we assumed that $\matr{P}$ is an $h \times h$-matrix.  It is important
to remark that for the algorithms in
Secs.~\ref{sec:affine:scaling:algorithm:1}
and~\ref{sec:primal:dual:interior:point:algorithms:1} the vector $\vect{u}$
usually does not have to be found perfectly. It is good enough to find an
approximation of $\vect{u}$ that is close enough to the correct $\vect{u}$.
(For more details, see e.g.~\cite[Ch.~9]{Bertsimas:Tsitsiklis:97:1}.)

Using a standard gradient-type algorithm to find $\vect{u}$ might work.
However, the matrix $\matr{P}$ is often ill-conditioned, i.e., the ratio of
the largest to the smallest eigenvalue can be quite big (especially towards
the final iterations), and so the convergence speed of a gradient-type
algorithm might suffer considerably.

Therefore, more sophisticated approaches are desirable. Such an approach is
the conjugate-gradient algorithm which was introduced by Hestenes and
Stiefel~\cite{Hestenes:Stiefel:52:1}. (See Shewchuk's
paper~\cite{Shewchuk:94:1} for a very readable introduction to this topic and
for some historical comments.) This method is especially attractive when
$\matr{P}$ is sparse or when $\matr{P}$ can be written as a product of sparse
matrices, the latter being the case for LP decoding of LDPC codes.

In the context of the affine-scaling algorithm, e.g.~Resende and
Veiga~\cite{Resende:Veiga:93:1} used the conjugate-gradient algorithm to
efficiently solve the relevant equation systems. They also studied the
behavior of the conjugate-gradient algorithm with different pre-conditioners.

A quite different, yet interesting variant to solve the minimization problem
in~\eqref{eq:unconstrained:qp:1} is by using graphical models. Namely, one can
represent the cost function in~\eqref{eq:unconstrained:qp:1} by an
\emph{additive} factor graph~\cite{Kschischang:Frey:Loeliger:01, Forney:01:1,
  Loeliger:04:1}.  Of course, there are a variety of factor graph
representations for the this cost function, however, probably the most
reasonable choice in the context of LP decoding is to choose the factor graph
that looks topologically like the factor graph that is usually used for
sum-product or min-sum algorithm decoding of LDPC codes. One can then try to
find the solution with the help of the min-sum algorithm.

[Equivalently, one can look at the maximization problem
\begin{align}
  \text{maximize}\quad
    &\exp
       \left(
         -
         \frac{1}{2}
           \vect{u}^\tr \matr{P} \vect{u}
         +
         \langle
           \vect{v}, \vect{u}
         \rangle
       \right)
         \label{eq:unconstrained:gaussian:1} \\
  \text{subj.\ to}\quad
    &\vu \in \R^h.
       \nonumber
\end{align}
Here the function to be optimized is proportional to a Gaussian density and
can be represented with a Gaussian factor
graph~\cite{Kschischang:Frey:Loeliger:01, Forney:01:1, Loeliger:04:1}. (Which
in contrast to the above factor graph is a \emph{multiplicative} factor
graph.) One can then try to find the solution with the help of the max-product
algorithm, which in the case of Gaussian graphical models is equivalent (up to
proportionality constants) to the sum-product algorithm.]

The reason for this being an interesting approach is that the behavior of the
min-sum algorithm applied to a quadratic-cost-function factor graphs is much
better understood than for other factor graphs. E.g., it is known that if the
algorithm converges then the solution vector is correct. Moreover, by now
there are also practically verifiable sufficient conditions for
convergence~\cite{Malioutov:Johnson:Willsky:06:1, Moallemi:VanRoy:06:1:subm,
  Moallemi:VanRoy:07:2:subm}. However, the quadratic-cost-function factor
graphs needed for the above problem are more general than the special class of
quadratic-cost-function factor graphs considered in the cited papers. Of
course, one could represent the cost function
in~\eqref{eq:unconstrained:gaussian:1} by a factor graph within this special
class (so that the above-mentioned results are applicable), however, and quite
interestingly, when this cost function is represented by a factor graph that
is not in this special class, then the convergence conditions seem to be
(judging from some empirical evidence) less stringent. In fact, we obtained
some very interesting behavior in the context of the short-step affine-scaling
algorithm where only one iteration of the min-sum algorithm was performed per
iteration of the affine-scaling algorithm. (The min-sum algorithm was
initialized with the messages obtained in the previous affine-scaling
algorithm iteration.)

\section{Other Simplifications}

\label{sec:other:simplifications:1}

\begin{figure}
  \begin{center}
    \subfigure[]{\epsfig{file=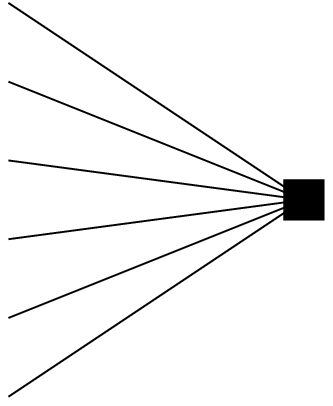, width=0.2\linewidth}}
    \hspace{2cm}
    \subfigure[]{\epsfig{file=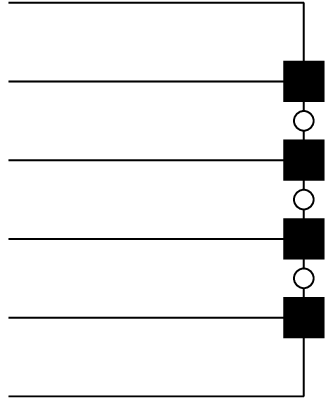, width=0.2\linewidth}}
  \end{center}
  \caption{Replacement of a partial factor graph representing a degree-$k$
    check function node by another partial factor graph with $k-2$ check
    nodes of degree three and with $k-3$ new auxiliary variable nodes. (Here
    $k = 6$.)}
  \label{fig:redrawing:check:function:node:1}
\end{figure}

Depending on the used algorithm, there are many small (but very useful)
variations that can lead -- when properly applied -- to considerable
simplifications. E.g., one can replace the partial factor graph in
Fig.~\ref{fig:redrawing:check:function:node:1}(a) by the partial factor graph
in Fig.~\ref{fig:redrawing:check:function:node:1}(b) that contains new
auxiliary variable nodes but contains only check nodes of degree
three~\cite{Chertkov:Stepanov:07:1, Yang:Wang:Feldman:07:1}. Or, one can
adaptively modify the set of inequalities that are included in the LP
formulation~\cite{Taghavi:Siegel:06:1, Wadayama:07:1:subm}.

\section{Conclusion}

\label{sec:conclusions:1}

We have presented some initial considerations towards using interior-point
algorithms for obtaining efficient LP decoders. Encouraging preliminary
results have been obtained but more research is needed to fully understand and
exploit the potential of these algorithms.

\section*{Acknowledgment}

This research was partly supported by NSF Grant CCF-0514801.

\end{document}